\documentclass[%
 pof,
 amsmath,amssymb,
 preprint,
 superscriptaddress,
 nofootinbib
]{revtex4-2}

\usepackage{graphicx}
\usepackage{dcolumn}
\usepackage{bm}
\usepackage[utf8]{inputenc}
\usepackage[T1]{fontenc}
\usepackage{mathptmx}
\usepackage{etoolbox}
\usepackage{booktabs}
\usepackage{gensymb}
\usepackage[colorlinks=true, allcolors=blue]{hyperref}
\usepackage[english]{babel}

\renewcommand{\selectlanguage}[1]{}

\newcommand\blfootnote[1]{%
  \begingroup
  \renewcommand\thefootnote{}\footnote{#1}%
  \addtocounter{footnote}{-1}%
  \endgroup
}

\makeatletter
\def\@email#1#2{%
 \endgroup
 \patchcmd{\titleblock@produce}
  {\frontmatter@RRAPformat}
  {\frontmatter@RRAPformat{\produce@RRAP{*#1\href{mailto:#2}{#2}}}\frontmatter@RRAPformat}
  {}{}
}%
\makeatother

\begin{document}

\title{Capillary force-driven particle orientation in rod networks}

\author{Lingyue Liu$^\ast$}
\affiliation{KU Leuven, Department of Chemical Engineering, Soft Matter, Rheology and Technology, 3001 Leuven, Belgium.}

\author{Sebastian Gassenmeier}
\affiliation{KU Leuven, Department of Chemical Engineering, Soft Matter, Rheology and Technology, 3001 Leuven, Belgium.}

\author{Erin Koos$^\ast$}
\affiliation{KU Leuven, Department of Chemical Engineering, Soft Matter, Rheology and Technology, 3001 Leuven, Belgium.}

\maketitle

\section*{Abstract}

Hypothesis: Anisotropic rod particles in capillary suspensions form complex network structures with distinctive orientation patterns and rheological properties that differ significantly from spherical particle systems. By identifying the orientation of individual particles, we are able to acquire invaluable experimental insight into the bulk particle orientation measurements.

Experiments: Glass microrods were dispersed in capillary suspensions with varied secondary liquid volume fractions. The resulting microstructural characteristics were analyzed using confocal microscopy. Meanwhile, their rheological properties were measured through rheometry and rheoconfocal techniques. Particle networks were quantified in terms of coordination number, clustering coefficient, and orientation distribution.

Findings: As the secondary liquid volume fraction increased, rod networks transitioned from point-to-point contact configurations to side-to-side aligned clusters. Unlike spherical systems, the average clustering coefficient decreased with increasing coordination number, indicating the formation of complex particle cluster configurations beyond simple side-to-side alignment. The rod networks demonstrated higher sensitivity to deformation, and samples with higher side-to-side contact probability exhibit higher viscoplastic fragility. These results provide a foundation for designing advanced materials with precisely tunable mechanical properties through controlled anisotropic particle interactions in capillary suspensions.

\blfootnote{\textit{$*$ Corresponding author: lingyue.liu@kuleuven.be ~~ erin.koos@kuleuven.be}}

\section*{Keywords}
Capillary suspension; Anisotropic particle; Rheology; Yielding; Network structure

\section*{Highlights}
\begin{enumerate}
\item	Rod-based capillary suspensions change contact type from point-to-point to side-to-side contact as secondary liquid increases.

\item	The clustering coefficient of rods decreases with increasing coordination number.

\item	Higher side-to-side contact probability increases viscoplastic fragility

\item   Particle clusters rotate and translate towards different directions during yielding.

\end{enumerate}

\newpage


\section{Introduction}

Anisotropic particles, such as plates~\cite{Ahn2016}, fibers~\cite{Ferec2009}, and rods~\cite{Solomon2010,Lewandowski2009}, exhibit distinct properties, including orientation-dependent interactions, enhanced percolation, and directional assembly. Moreover, these particles can introduce unique functionalities, such as electrical or thermal conductivity, optical anisotropy, and mechanical reinforcement, which are highly desirable in various industrial applications~\cite{Lee2011}. The common spherical models fail to describe rod particle suspensions because their elongated geometry creates orientation-dependent hydrodynamic forces and rotational dynamics. Different from spherical particles that rotate freely, anisotropic particles tumble and exhibit transient properties depending on their orientation~\cite{Mueller2009}. In semi-dilute suspensions, particle-particle contact interactions contribute to the viscosity with a quadratic dependency on particle volume fraction~\cite{Ferec2009}. As the volume fraction keeps increasing, the minimum separation between the nearest neighboring particles decreases, and a yield stress is developed when they are in repulsion, forming a jammed network. The network is capable of elastically accommodating stress until the yield stress is reached and subsequent breakdown and shear-thinning behavior is observed~\cite{Khan2023}. Using particles with higher aspect ratios, the maximum random close packing and the percolation threshold are reduced, allowing networks to form at lower particle density~\cite{Khan2023,Mueller2009}.

The percolating network formed using anisotropic particles exhibits complex microstructural rearrangements and distinctive orientation behavior that significantly influence their rheological properties under external forces. Most importantly, particle shape determines the material response: discoid particles dominated by face-to-face contacts with high contact areas exhibit higher elasticity~\cite{Kao2022}, whereas rod gels tend to weaken after preshear due to cluster densification and increased structural heterogeneity, particularly under oscillatory shear conditions~\cite{Das2022}. Further, rod suspensions can form distinctive flow-aligned~\cite{Calabrese2023} or
vorticity-aligned structures~\cite{Das2021,Das2022}, and may eventually break and induce two-step yielding at higher shear rates~\cite{Shakeel2021}. The response depends on the particle flexibility~\cite{Lang2019}, viscous drag~\cite{Das2022}, 
interparticle interactions~\cite{Calabrese2023,Lang2019, Kao2022}, and thermal motion~\cite{Dhont2006,Calabrese2023}. In this work, we present the effect of capillary force provided by interparticle liquid bridges that are a few magnitudes higher than the aforementioned forces, forming a novel class of viscoelastic material named capillary suspension~\cite{Koos2011}.

Capillary suspensions, a class of materials formed by adding a small amount of immiscible secondary fluid to a normal suspension, have attracted attention in recent years due to their unique rheological properties. From an industrial perspective, capillary suspensions offer a promising route to formulate artworks~\cite{Ranquet2023}, porous ceramics~\cite{Menne2022}, or precursors for 3D printing~\cite{Weis2020,Liu2025}. The added secondary fluid forms a sample-spanning network of bridged particles, which drastically changes the mechanical properties by introducing a yield stress at a much lower particle volume fraction in comparison to the network percolating threshold~\cite{Koos2011}. While the majority of the studies on capillary suspensions have been focusing on spherical particles~\cite{Allard2022, Bossler2017, Bossler2018, Natalia2022}, there are certain ones with anisotropic (non-arbitrarily-shaped) microparticles, with oblate~\cite{Kazama2022,Maurath2016} and prolate particles~\cite{Maurath2016,Qiao2019}. The incorporation of anisotropic particles can extend the range of achievable properties and enable the development of novel, high-performance materials. 

The local microstructure of capillary suspensions, which governs the system, can be tuned by changing several key factors, particularly particle shape~\cite{Maurath2016}, aspect ratio~\cite{Mueller2009}, and secondary fluid volume fraction~\cite{Bindgen2020}. At small fractions of secondary liquid, increasing the liquid volume fractions leads to a higher probability of bridge formation between adjacent particles, with larger volume fractions resulting in more robust and stable bridges. As the amount of secondary liquid increases, both coordination number and clustering coefficient increase positively, reflecting the structural transitions from pendular to funicular and further to capillary states~\cite{Bindgen2020,Bindgen2022}. This effect is particularly pronounced in anisotropic particle networks, where the interplay between aspect ratio and secondary fluid creates network structures whose formation mechanisms and topological properties require systematic investigation~\cite{Maurath2016}. Within the networks, the precise orientation of individual particles, the local cluster configuration, and the internal structures' response to external deformation remain unknown.

In this context, the present paper aims to address fundamental questions about capillary suspensions containing rod-like particles, investigating the influence of the capillary interaction on particle orientation, network formation, and structural stability. The response of these systems to external stresses is explored using a combined rheometer and confocal microscope. Here, we aim to establish a comprehensive understanding of these complex systems and develop principles for designing and optimizing capillary suspensions for targeted performance.


\section{Materials and methods}

\subsection{Sample preparation}

Glass microrods (PF-60S, Nippon Electric Glass Co., Ltd.) with an average diameter of 6~$\pm$~0.01~$\mathrm{\mu}$m and an aspect ratio of 4~$\pm$~1 were used as the main structure of the capillary suspensions. The microrods, with a density of 2.6~g/cm$^3$ and a refractive index of n = 1.56, were fluorescently dyed with rhodamine B isothiocyanate (RBITC, Sigma-Aldrich). The bulk phase of the capillary suspensions consisted of a mixture of 82.2~wt\% cinnamon bark oil ($>$ 99\%, Sigma-Aldrich) and 17.8~wt\% 1,2-cyclohexane dicarboxylic acid diisononyl ester (pure, Hexamoll DINCH, BASF). The oil mixture had a refractive index of $n=1.56$, matching that of the microrods. The viscosity of the oil mixture is 7.0~mPa$\cdot$s. The density of this mixture is 1.01~g/ml. The secondary liquid was a mixture of 50~vol\% glycerol ($>$ 99.5\% , Carl Roth) and 50 vol\% ultrapure water (Arium 611DI, Sartorius Stedim Biotech) with a $n=1.4$ and a density of 1.12~g/ml, dyed with PromoFluor 488 premium carboxylic acid (PromoCell GmbH). While this solution did not match the refractive index of the particles and bulk liquid, the percentage of glycerol was kept low to prevent an elevated viscosity, which would require excessive energy input during dispersion. High energy input during dispersion of the rods and breakup of the secondary liquid droplets accelerated cinnamon oil oxidation, which modified the bulk liquid refractive index, minimizing the visual depth. 

156~mg (15~vol\%) rods were dispersed in the 337~--~338.5~$\mathrm{\mu}$l (84.25~--~85 ~vol\%) bulk liquid to ensure homogeneous wetting and prevent aggregation; the suspension has a viscosity of 22.0~mPa$\cdot$s. The corresponding volumes of secondary liquid 0~--~3~$\mathrm{\mu}$l (0~--~0.75~vol\%) were then dispersed using an ultrasonic horn of 3.175~mm diameter (Digital Sonifier model 450, Branson Ultrasonics corporation) at a 10\%  amplitude for 2~s. 
The interfacial tension between the secondary liquid and the bulk liquid mixture is 11.7 $\pm$ 0.2~mN/m, as measured with a drop tensiometer (Attension, Biolin Scientific) at room temperature (20.5 $\pm$ 1$^{\degree}$C) using a frame rate of 20 FPS and a duration of 30~min. The value did not fluctuate throughout the experiment period. The three-phase contact angle $\theta$ of the rods was determined using confocal images of the particles at the liquid-liquid interface in a microchannel, following the method of Allard et~al.~\cite{Allard2022}. This contact angle confirms the strong hydrophilic nature of the rods, which are completely immersed (0\textdegree) within the water phase (supplementary Fig.~S1).

\subsection{Rheological measurements}

Due to the limited sample quantity, a stress-controlled rheometer MCR702 (Anton Paar) with a plate-plate geometry of 8~mm diameter (PP08) was used to perform oscillatory measurements. The TwinDrive rheometer was measured in separate motor transducer mode for enhanced sensitivity. All amplitude sweeps were conducted at an angular frequency of 10 rad/s. Frequency sweeps with a shear strain of 0.005\%  ensured the frequency independence of the gel structures between 0.1 rad/s and 10 rad/s, as shown in the supplemental data Fig.~S2, with solid-like behavior ($G^{\prime}$ > $G^{\prime\prime}$). The relaxation time is much longer than the experiment time (when excluding factors such as evaporation) as the capillary force between particles is much stronger than factors such as the thermal fluctuation.
Viscosity measurements for bulk liquid and suspension without secondary liquid were performed from 0.1~to 10~s$^{-1}$ with a duration of 10~s per point, with a top geometry of PP25 to ensure trustworthy measurement.

The sample was loaded onto the rheometer with a spatula. The top plate was lowered to a gap of 3 mm, after which the compression was recorded. The top plate was then lowered to the trim position of 1.025~mm, 25~$\mathrm{\mu}$m above the measuring position, with an uniaxial velocity of 19.75~$\mathrm{\mu}$m/s. This allowed the extra sample at the edges to be scraped off prior to other measurements. Afterward, the top plate was lowered to the measurement gap of 1 mm. The normal force, i.e., the thrust on the plate, was recorded during both the gap-lowering and oscillating experiments.

\subsection{Confocal and rheoconfocal microscopy}

Samples were spread onto a 22~$\times$~40 mm cover glass (Menzel-Gl\"{a}ser) using a spatula. The samples were then transferred to a Leica TCS SP8 inverted confocal laser scanning microscope (Leica Microsystems GmbH, Wetzlar, Germany). Imaging of the capillary suspension microstructure was performed using a glycerol immersion objective with 63$\times$ magnification and a numerical aperture of 1.3. Solid-state lasers with wavelengths of 488 nm and 552 nm were used to excite PromoFluor 488 premium carboxylic acid (liquid channel) and rhodamine B isothiocyanate (particle channel), respectively. To reconstruct the 3D microstructure of the capillary suspension, 3D image stacks were taken, each consisting of $x$-$y$ planes of size 246 $\times$ 246~$\mathrm{\mu}$m and a resolution of 1024 $\times$ 1024 pixels. The $z$ resolution was set to 3 micrographs per 1~$\mathrm{\mu}$m, the intensity loss over $z$ due to the high refractive index of the system allowed 210 micrographs, with an equivalent of 70~$\mathrm{\mu}$m, to be taken before micrographs become inaccessible for further analysis. 

To observe sample deformation under shear, a stress-controlled rheometer (MCR302 WSP, Anton Paar) was mounted on a microscope stand (IX71, Olympus) equipped with a fast-scanning confocal microscope (VT HAWK, Visitech) to perform simultaneous imaging and rheological measurements. A stiff sapphire glass window with a diameter of 30 mm and a thickness of 170~$\mathrm{\mu}$m was glued into a custom-made holder as the bottom geometry. A Hamamatsu EMCCD camera recorded image sequences with exposure times of 50 ms in 2D and 100 ms in 3D. Imaging was done using two oil objectives (UPlanSApo, Olympus) with magnifications of 20$\times$ and 60$\times$ and numerical apertures of 0.85 and 1.35, resulting in pixel sizes of 0.80~$\mathrm{\mu}$m and 0.27~$\mathrm{\mu}$m, respectively. The objectives were mounted on a piezo actuator to perform 3D scans in the $z$-direction. The rheometer was positioned on an $xy$-gantry system (Screw Drive LRT Gantry, Zaber) to reposition it relative to the microscope, allowing measurements at different radial positions along the 15 mm diameter stainless steel parallel plate geometry. The imaging plane (horizontal) was positioned 1 mm from the edge of the plate. The rheometer applied the input strain at 2/3 of the plate radius, with calculation mode DIN 53018 R2/3, allowing local strain to be calculated by $\gamma_{test,0} \cdot \frac{3}{2} \cdot \frac{6.5\mathrm{\ mm}}{7.5\mathrm{\ mm}}$. Rheological measurements were performed at an angular frequency of 10 rad/s with gaps between 200~$\mathrm{\mu}$m and 300~$\mathrm{\mu}$m, minimizing the effect of potential confinement. A doctor blade was used to load thin capillary suspension layers, avoiding solid phase compaction due to such small gaps.

\subsection{Image processing and particle tracking}

\begin{figure}[tbp]
    \centering
      \includegraphics[width=0.7\textwidth]{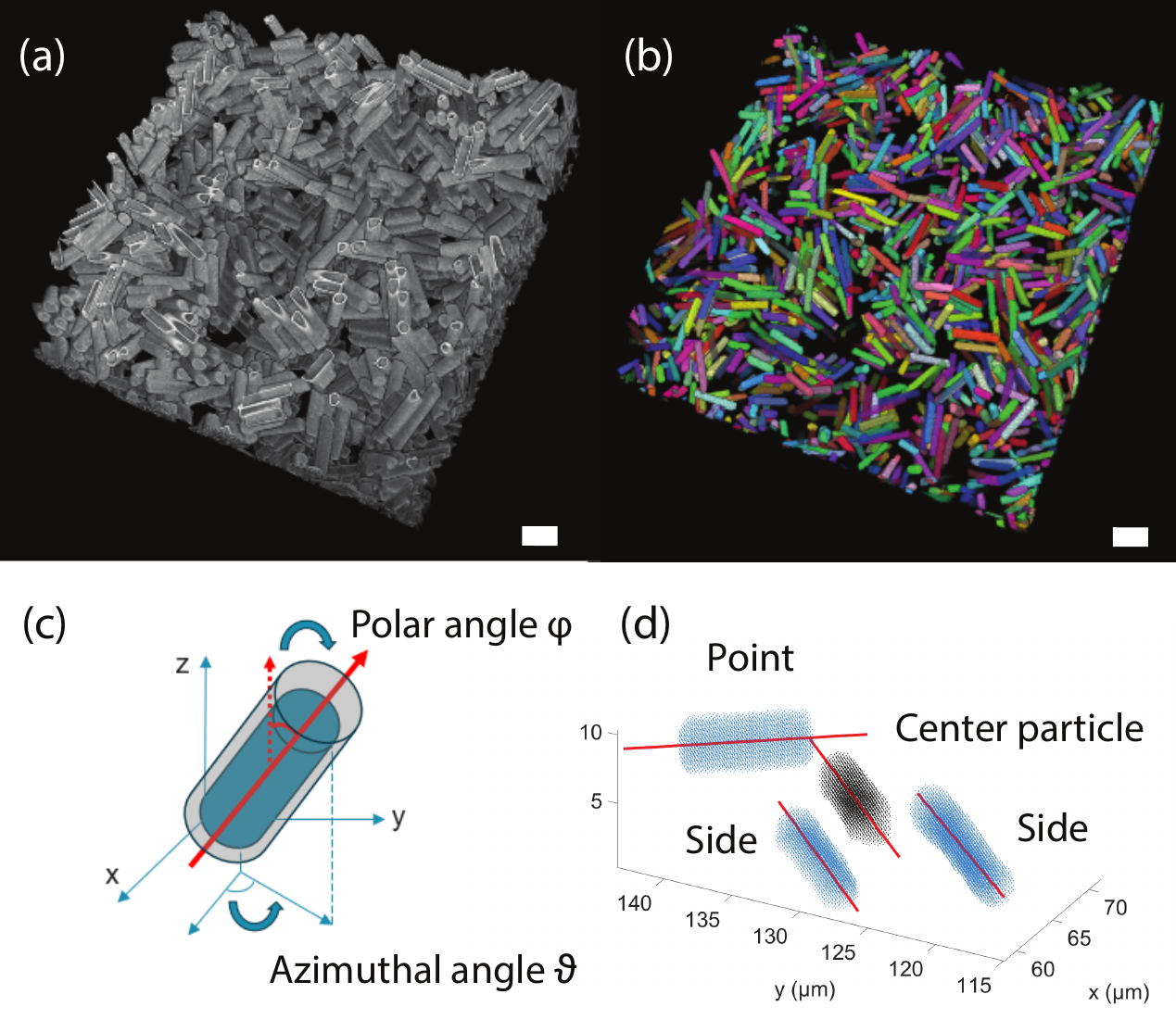}
      \caption[3D microstructure imaging and orientation analysis of rod-based capillary suspensions.]{(a) 3D scanned micrographs of a capillary suspension gel with 15~vol\% of 6~$\mathrm{\mu}$m rods with 0.5625\%  secondary liquid (b) detected and reconstructed 3D view of the sample with each particle colored differently, the scale bars are 20~$\mathrm{\mu}$m. (c) Schematic representation of rod orientation resolved by azimuthal angle $\vartheta$ and polar angle $\varphi$ (d) Example of particle contact classification, where the black voxels are the center particle and blue voxels are neighboring particles. Red lines indicate calculated major axes.. The particles appear not to be in direct contact with each other due to the exclusion of their fluorescent outlines.}
      \label{fig:3D}
    \end{figure}

The confocal micrographs of the static structure were exported as two separate channels for fluorescence response of the two lasers illuminating the microrods and secondary liquid. The microparticle channel (Fig.~\ref{fig:3D}a) 
was denoised in Fiji using the subtract background function with a rolling ball radius of 50 pixels. Next, we auto-local thresholded each micrograph using the Niblack method with a 15-pixel radius to sharpen and identify edges. This process allowed for the separate detection of enclosed areas based on their fluorescent boundaries. We then tracked and linked these enclosed areas (Fig.~\ref{fig:3D}b) using centroid distance criteria through a modified MATLAB 3D tracking code, adapted from the algorithm developed by E.R.~Weeks and J.C.~Crocker~\cite{Weeks2000}. The detected image appears sparser in comparison to the original image due to the fact that the shells are discarded and the particles are reconstructed using the enclosed center areas within the fluorescent outlines. These outer shell signals have an approximate length of 0.6~$\mathrm{\mu}$m, resulting in a lower apparent solid volume fraction.

We resolve the orientation of the individual rods, represented by their respective voxel lists, by solving the eigenvalue problem of the moment of inertia matrix $I_{ij}=m_\alpha r_\alpha^2 \delta_{ij}-m_\alpha r_{\alpha i} r_{\alpha j}$ for each rod, which is symmetric and therefore yields positive eigenvalues and an orthonormal eigenbasis. We are then able to discriminate the moment of inertia in the longitudinal direction of the rod and set the corresponding eigenvector as the rod orientation. This orientation is transformed to spherical coordinates, yielding a polar angle $\varphi$ ($0 \leq \vartheta \leq \frac{\pi}{2}$) as the angle formed by the positive z-axis (polar axis) and the radial line and the azimuthal angle $\vartheta$ ($0 \leq \varphi \leq \pi$) which is the angle of rotation around the z-axis (Fig.~\ref{fig:3D}c).

To quantify the network structure, two general parameters are introduced: the coordination number ($z$) and the clustering coefficient ($c$). The coordination number is the number of neighboring particles that are in contact with each particle. Based on the averaged outline thickness of the detected fluorescent shell, a 3~$\mathrm{\mu}$m proximity criterion was established for contact determination, accounting for the spatial resolution limitations of the confocal microscopy system. Unlike a spherical system with a maximal coordination number of 12~\cite{Bindgen2020}, this maximal value is related to the average aspect ratio of the applied rods. The clustering coefficient is an indication of the degree of local aggregation, defined as $c =\frac{2e}{z(z-1)}$, where $e$ is the number of connections between the neighboring particles~\cite{Bindgen2020}. The clustering coefficient $c$ therefore divides the neighbor-connectivity by the maximum possible connectivity and yields values from 0 to 1 for all systems, denoting a loosely interconnected or a fully clustered network, respectively. 

To further quantify the morphology of the system, the particle contacts can be categorized into two types, point-to-point or side-to-side, as shown in Fig.~\ref{fig:3D}d). Due to the unevenness of the rod particle ends, end-to-end contacts are not statistically significant in this study. As a general parameter, we introduce the side contact ratio since side-to-side contact pairs exhibit a much stronger capillary force due to the elevated contact area in comparison to the point-to-point contact pairs. The enhanced capillary forces promote aggregation, resulting in a rod cluster self-aligning in the same direction. Therefore, the criteria for side-to-side contact area whether the touching particles are aligning in the same direction ({$\pm$~10\degree} for both $\varphi$ and $\varphi$, Fig.~\ref{fig:3D}d).


\section{Results and discussion}

\subsection{Microstructure}

\begin{figure}[tbp]
    \centering
      \includegraphics[width=0.7\textwidth]{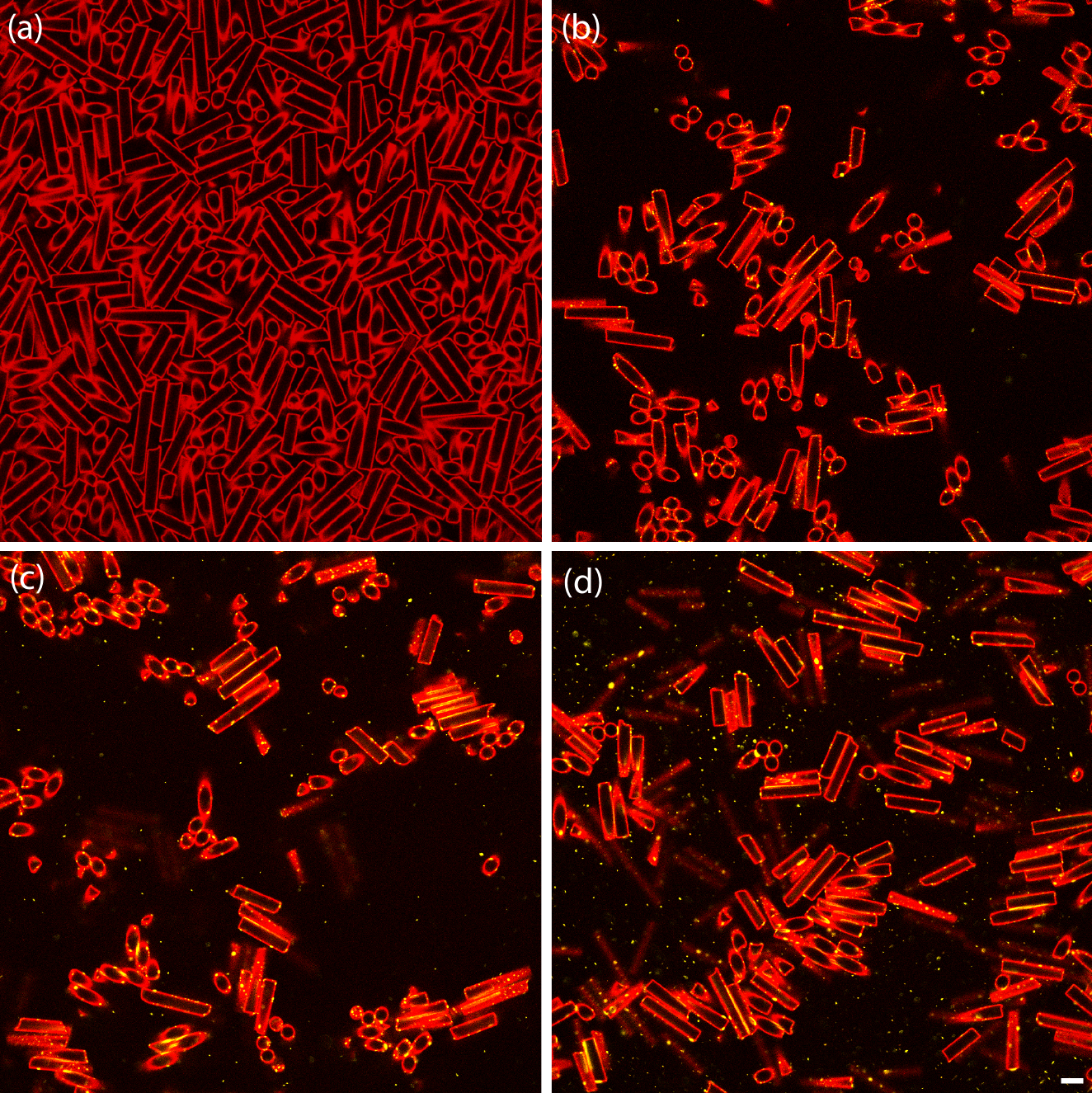}
      \caption[Confocal micrographs of the evolution of microstructure with increasing secondary liquid content.]{Confocal micrographs of capillary suspension with different amount of secondary liquid (a) 0\%  ($\phi_\mathrm{sec}/\phi_\mathrm{solid}$ = 0.0\% ) (b) 0.375\%  ($\phi_\mathrm{sec}/\phi_\mathrm{solid}$ = 2.5\% ) (c) 0.5625\%  ($\phi_\mathrm{sec}/\phi_\mathrm{solid}$ = 3.75\% ) (d) 0.75\%  ($\phi_\mathrm{sec}/\phi_\mathrm{solid}$ = 5\% ). Red regions show rod particles (shells), while yellow areas represent capillary bridges, and the dark background is the bulk liquid. The scale bar is 10~$\mathrm{\mu}$m. }
      \label{fig:confocal}
    \end{figure}

Example micrographs of the network captured via confocal microscopy are shown in Fig.~\ref{fig:confocal}. 
The micrographs, showing slices near the bottom plate, reveal a pronounced evolution in microstructure with increasing secondary fluid content. In Fig.~\ref{fig:confocal}a, the suspension system shows a relatively uniform, compact distribution of rod particles that sediment due to their weight and lack of interparticle forces. The majority of the rods lay flat (parallel to the image plane and perpendicular to gravity) with a polar angle approaching {90\textdegree}. The maximum packing $\phi_\mathrm{m,rod}$ of rods with an aspect ratio $AR$ is given by  Eq.~\ref{eq:maxPacking}~\cite{Mueller2009},
    \begin{equation}
    \phi_\mathrm{m,rod} = \frac{2}{0.321AR + 3.02}
    \label{eq:maxPacking}
    \end{equation}
For rod-shaped particles of AR = 4~$\pm$~1, the maximum packing  decreases slightly from $\phi_{m,rod}=64$\%  for spheres to $\phi_\mathrm{m,sph}=46.5$~vol\%~\cite{Li2010}. The effective volume fraction calculated for the sedimented rods in Fig.~\ref{fig:confocal}a is 44.1\%, close to the theoretical value. However, due to the rod length distribution, the real maximum packing should be higher than the theoretical value~\cite{Lam1998}.  A marked difference is observed with the addition of secondary fluid, as shown in Fig.~\ref{fig:confocal}b-c, where a gel-forming structure (characterized by a sample-spanning network connected by capillary bridges) is observed with capillary bridges shown in yellow. The gel network forms at 15~vol\% rods, a concentration much lower than the estimated percolation threshold of 0.8$ \phi_\mathrm{m,rod} =37.2$~vol\% Therefore, the capillary force stemming from surface tension and capillary pressure gives rise to a gel structure in these capillary suspension samples~\cite{Princen1968,Princen1969,Princen1970}.   

As the secondary fluid content increases, the particles aggregate progressively, characterized by clusters of rods with preferential side-to-side alignment, as shown in Supplementary Fig.~S3. This clustering behavior is attributed to the capillary force increase by introducing excessive secondary fluid, yielding a transition in preferred configuration from point-to-point to side-to-side contacts. The point contact capillary force is estimated using the model developed by Rabinovich~et~al., as shown in Eq.~\ref{eq:capFsphere}~\cite{Rabinovich2005},
    \begin{equation}
    \begin{aligned}\label{eq:capFsphere}
    F_\mathrm{cap, point} &= \frac{2\pi r \gamma  \cos\theta}{1+\frac{S}{2C}} \\
    \mathrm{for} \quad C &=\frac{S}{2}\left(-1+\sqrt{1+ \frac{2V}{\pi r S^2}}\right) \\
    \end{aligned}    
    \end{equation}
where $\gamma$ is the interfacial tension, the three-phase contact angle is $\theta$, $r$ is the radius of the rod, $V$ is the individual bridge volume and $S$ is the separation distance. Using a cylindrical approximation to find the bridge volume, the corresponding capillary force is approximately 200~nN. 

In wet spherical granular systems, bridges are generally toroidal, with the possibility of forming trimer, pentamer, or filled tetrahedra~\cite{Scheel2008}. In rod systems, however, the formation of larger bridges is observed less frequently when the secondary liquid preferably wets the solid particles~\cite{Princen1969}. This is because the filled bridges are not thermodynamically stable and tend to separate when particles are in close contact in a side-to-side structure

This occurs because the secondary liquid minimizes its surface energy by forming separate bridges rather than large filled structures when rods are in close side-to-side contact~\cite{Princen1969}. Based on the critical separation model proposed for the three-cylinder system~\cite{Princen1970}, as long as the separation distance $S_\mathrm{crit}/r < 0.28$, the capillary bridges remain separate for $\theta$ = 30$^{\degree}$ rods, as shown in Supplementary Fig.~S4b. Above this distance, the thermally equilibrated stable bridges with high menisci merge again and fill the internal gap~\cite{Princen1970,Princen1968,Princen1969}. The critical separation distance in the system applied in this study is 0.8~$\mathrm{\mu}$m, exceeding the maximum bridge height, the separation of the bridge is confirmed through the observation of the micrograph. 

Therefore, the toroidal bridge between two rods stretches and expands over the whole length of the rods, and the capillary force per unit length $F_\mathrm{cap, side}/L$ is given in Eq.~\ref{eq:capFperunit}~\cite{Princen1970},
    \begin{equation}
        F_\mathrm{cap, side}/L= 2\gamma~\sin(\alpha + \theta) + 2\gamma~\sin\alpha(r /R) 
    \label{eq:capFperunit}
    \end{equation}
where the angle between the line connecting the centers of the cylinders and the radius to the liquid-liquid boundary is $\alpha$,  and $R$ is the curvature of the liquid surface, as shown in the schematic shown in Fig.~S4a. By manually categorizing over 20 samples for each contact type (point-to-point contacts and side-to-side contact) in the 3D stacks of Fig.~\ref{fig:confocal}d (Supplementary Fig.~S5), the observed average bridge height is approximately 0.6 $\pm$ 0.1~$\mathrm{\mu}$m, and the averaged width of the bridges is to 2.4 $\pm$ 0.1~$\mathrm{\mu}$m, despite their high variability in bridge length. Taking these values into account, the calculated angle $\alpha$ = 25\textdegree, the contact angle $\theta$ is 28\textdegree, which shows their strong hydrophilic nature but slightly higher than the value measured using the microchannel liquid-liquid interface contact angle technique (0\textdegree) shown in supplementary Fig.~S1.

Taking these values into account, the calculated angle $\alpha$ = 25\textdegree, and the contact angle $\theta$ = 28\textdegree. This contact angle confirms the strong hydrophilic nature of the rods, though it is slightly higher than the value of 0\textdegree measured using the microchannel liquid-liquid interface technique (supplementary Fig.~S1).

By applying the $\theta$ and $\alpha$ into Eq.~\ref{eq:capFperunit}, $r/R \approx$ 5.97 and the resulting capillary force per unit length  is 0.078~N/m. Taking the average rod length ($L$) into consideration, the total capillary force is roughly 1.87~$\mathrm{\mu}$N. Given these values, the ratio between $F_\mathrm{cap,side}$ and $F_\mathrm{cap,point}$ is up to 935\%, which promotes side-to-side particle attraction. 
Regarding the role of particle weight, if we consider a cluster of rods suspended from a single bridge, gravity only overcomes the capillary force for 1.9*10$^5$ rods (side-to-side) or 2*10$^4$ (point-to-point). These values indicate that gravitation can be ignored as the particle number in the field of view is in the range of 10$^3$. To better assess the influence of the secondary liquid content, we can calculate the particle orientation. 

\subsubsection{Polar angle and azimuthal angle}

\begin{figure}[tbp]
    \centering
      \includegraphics[width=0.93\textwidth]{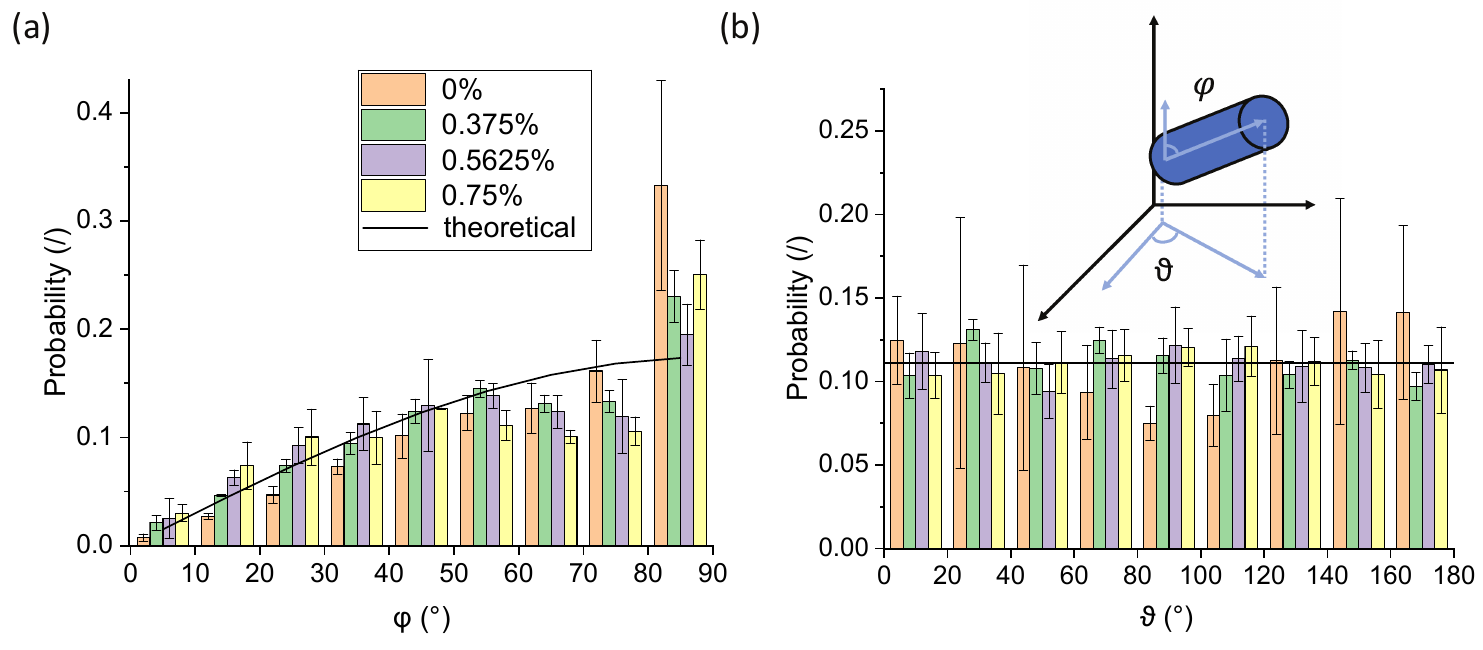}
      \caption[polar angle and azimuthal angle distribution of rod particles in capillary suspensions.]{Probability distribution of (a) polar angles ($\varphi$), ranging from 0$^{\degree}$ (perpendicular to the substrate) to 90$^{\degree}$ (parallel to the substrate) and (b) azimuthal angles ($\vartheta$), ranging from 0$^{\degree}$ to 180$^{\degree}$. A simplified sketch of rod orientation is attached. Different colors represent different secondary liquid fractions: orange (0\%), green (0.375\%), purple (0.5625\%), and yellow (0.75\%). The black curves show the theoretical probability for an isotropic system. }
      \label{fig:TwoAngles}
    \end{figure}

With increasing secondary liquid volume fractions, the alignment of the particles with the glass plate (parallel to the image plane in Fig.~\ref{fig:confocal}) decreases. This phenomenon is particularly pronounced for 0.375~vol\% secondary liquid, shown in Fig.~\ref{fig:confocal}b, where the majority of particles do not align parallel to the glass plate. The polar angle probability distribution is as shown in Fig.~\ref{fig:TwoAngles}a, 
where the theoretical probability of finding a particle with polar angle $\varphi$ in an isotropic system within the range $[\varphi_{1}, \varphi_{2}]$ is shown using the black curve. This probability can be expressed as,
    \begin{equation}
    \mathbb {P} (\varphi|_1^2) = \frac{A (\varphi|_1^2)}{A|_0^\frac{\pi}{2}} = \frac{\int_0^{2\pi} \int_{\varphi_{1}}^{\varphi_{2}} L^2 \sin\varphi \, d\varphi \, d\vartheta}{2\pi L^2}
    \end{equation}
where the particle length $L$ is the radial distance in the spherical coordinate system. Without secondary liquid, the polar angle distribution $\varphi > 80^{\degree}$ indicates that the majority of the rods are lying against the substrate due to gravity, deviating from the theoretical value. The capillary force between particles becomes effective as secondary liquid bridges are formed, as shown by the decrease in possibility in higher polar angles and increase in the probability at lower angles. As the liquid volume fraction increases, the probability distribution of the polar angles at lower angles increases, approaching the theoretical value. This behavior can be attributed to the elevated point-to-point interparticle contacts. However, as the liquid volume fraction approaches 0.75\%, the probability of finding particles located at polar angles $80^{\degree} < \phi < 90^{\degree}$ increases. This occurs because the elevated liquid level enables much more probable side-to-side contact between particles, especially when particles face a flat substrate (bottom plate) rather than highly curved interparticle surfaces. When in contact with both configurations, the orientation competition lies between side-to-side (particle/cluster-glass) and point-to-point (particle/cluster-particle) capillary bridges. The side-to-side contacts with high capillary forces may result in preferential high polar angles, as can be seen in Supplementary Fig.~S6d. This biased structure approaching the bottom slide is not reported in spherical systems where only point-to-point contact exists~\cite{Liu2024,Allard2022}.

The azimuthal angle (Fig.~\ref{fig:TwoAngles}b) does not seem to be affected by the increasing amount of secondary liquids, following the theoretical line for a random orientation. The only exception to this homogeneous distribution is the 0\% sample where the capillary force is absent. The suspension system exhibits larger error bars and a decrease in the probability of finding rods with $80^{\degree} < \vartheta < 120^{\degree}$, indicating substantial unity in particle orientations. Such a bias in the azimuthal angle may be due to the influence of sample loading (e.g., strain caused by the spatula or tilting of the slide).

\subsubsection{Coordination number and clustering coefficient}

\begin{figure}[tbp]
    \centering
      \includegraphics[width=0.8\textwidth]{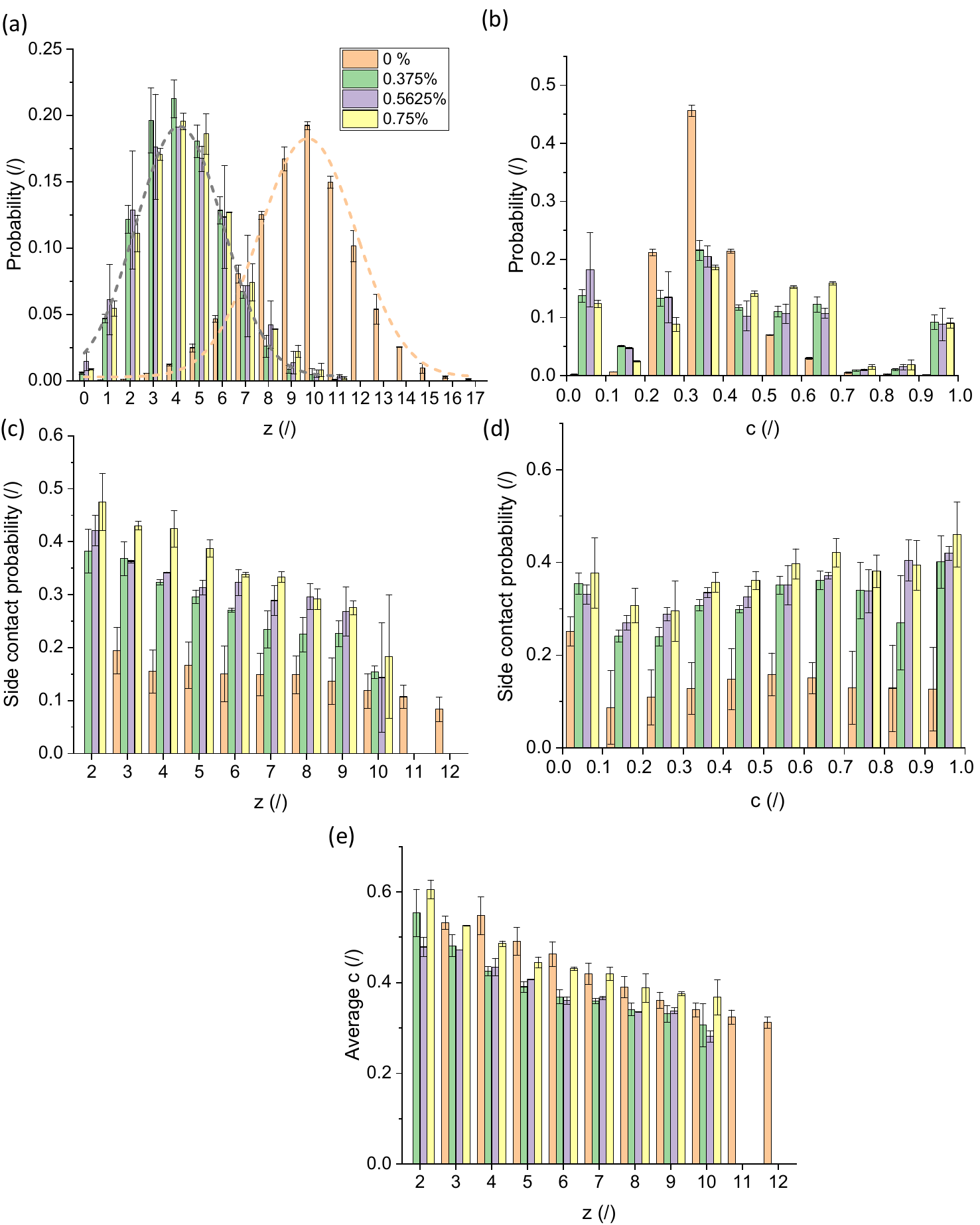}
      \caption{Probability distribution of average (a) coordination number ($z$) and (b) clustering coefficient ($c$) with different amounts of secondary liquids shown in different colors representing different secondary liquid fractions: orange (0\% ), green (0.375\% ), purple (0.5625\% ), and yellow (0.75\% ). Average side contact probability distribution over (c) coordination number and (d) clustering coefficient with different amounts of secondary liquids. (e) Average clustering coefficient over coordination number. The gray and orange dashed lines in (a) show the fits of two normal distribution curves.}
      \label{fig:microstructure}
    \end{figure}

Since there are clear differences in the microstructure resulting in the apparent bias in polar angle, we also quantified the coordination number and clustering coefficients~\cite{Bindgen2020}, as shown in Fig.~\ref{fig:microstructure}a 
and \ref{fig:microstructure}b respectively. 
In Fig.~\ref{fig:microstructure}a, the coordination number distribution shows that the suspension system without secondary fluid ($\phi_\mathrm{sec} = 0\%$) has a broad distribution $z$ (8 -- 12). The coordination numbers were fitted with a normal distribution having a mean value of 9.74 and a standard deviation of 2.11 (Full Width at Half Maximum, $\mathrm{FWHM}=4.96$). These values are consistent with a compact contact network, mirroring that typically found in granular media ($z = 9$--10)~\cite{Wouterse2009}. In contrast, the systems with added secondary fluid predominantly exhibit lower coordination numbers ($z = 2$--5). The decreased number in individual neighboring particles denotes a higher effective volume, resulting in a space-spanning network~\cite{Bindgen2020}. The three capillary suspension distributions show similar patterns with slightly different mean values of 4.05 for $\phi_\mathrm{sec} = 0.375\%$, 4.14 for $\phi_\mathrm{sec} = 0.5625\%$, and 4.20 for $\phi_\mathrm{sec} = 0.75\%$, with an average standard deviation of $1.90 \pm 0.04 $ and $\mathrm{FWHM}=4.48\pm0.09$ (fitted with gray dashed lines). Despite the apparent differences in structure shown in Fig.~\ref{fig:confocal}b to ~\ref{fig:confocal}d with $\phi_\mathrm{sec} = 0.375\%$ having smaller clusters and $\phi_\mathrm{sec} = 0.75\%$ having larger, the coordination number distributions show minimal differences, indicating that particles reorganize their orientation and contact type without significantly altering the number of contacts.

The analysis of particle contact topology reveals distinct trends in the microstructural organization of rod-based capillary suspension. Along with the averaged clustering coefficient, they are plotted over the coordination number, as shown in Fig.~\ref{fig:microstructure}c and~\ref{fig:microstructure}e. Particles with coordination numbers $z<2$ for suspension systems and $z<1$ for capillary suspension systems are primarily located at the edges of image stacks and are therefore excluded. Meanwhile, the particles with coordination numbers $z>12$ for suspension systems and  $z>10$ for capillary suspension systems are not statistically significant given their small number. Therefore, both cases are eliminated from the figures. The side-to-side contact probability exhibits a monotonic decrease with increasing coordination number, with maximal values observed at low coordination numbers ($z = 2$--4). The introduction of secondary fluid significantly enhances the probability of side-to-side contacts compared to the suspension system as capillary forces become significant (Fig.~\ref{fig:microstructure}d).

The clustering coefficient distribution in Fig.~\ref{fig:microstructure}b demonstrates a particularly special behavior where the suspension system ($\phi_\mathrm{sec} = 0\%$) shows a pronounced peak at higher clustering coefficients ($c\approx 0.3$) and a quasi-normal distribution (except for $c < 0.2$). Meanwhile, systems with secondary fluid display more uniform distributions across middle clustering coefficient values ($c = 0.2 $-- 0.7). This suggests that while secondary fluid promotes stable contacts between particles, it leads to more diverse local structural arrangements (see Fig.~\ref{fig:microstructure}d for examples) rather than the more uniformly clustered structures seen in the suspension system. 

The clustering coefficient distribution reveals notable characteristics at the extremes, specifically at $c = 0$~and~$c = 1$, representing fully sparsely connected and clustered configurations, respectively. Mathematically, a low (but non-zero) $c$ typically arises from structures with high $z$ with few inter-neighbor clusters, while $c = 0$ does not restrict the number of neighboring particles. The peak at $c = 0$ can either arise via a sparse network of point-to-point contacts without trimers, or clusters of side-to-side contacts without any inter-connections (Supplementary Fig.~S7). The average clustering coefficient decreases monotonically with coordination number across all secondary fluid volume fractions (Fig.~\ref{fig:microstructure}e). While secondary fluid content shows limited impact on the averaged clustering behavior, the inverse relationship between clustering coefficient and coordination number suggests a fundamental transition in network topology: low coordination configurations ($z < 4$) display high clustering coefficients ($c \approx 0.5$), indicating locally dense arrangements, while higher coordination numbers ($z > 8$) correspond to more distributed structures with clustering coefficients $c < 0.4$. 

As shown in the plot of side contact probability as a function of the clustering coefficient in Fig.~\ref{fig:microstructure}d, the percentage of side-to-side contacts generally increases with the clustering coefficient for the capillary suspensions, but there is a local maximum for $c < 0.1$. For the pure suspension, the side contact probability is significantly higher for $c < 0.1$, implying a flat structure without interconnects, likely caused by the sedimentation against the bottom plate, as shown in Fig.~\ref{fig:confocal}.

For particles with more neighboring particles (higher $z$), the network configuration becomes more difficult to fully cluster ($c = 1$)~\cite{Torquato2010}. As can be seen from Supplementary Fig.~S7, for a particle with $z = 6$, the theoretical maximum clustering coefficient for rods arranged in a hexagonal configuration with solely side-to-side contacts is limited to $c = 2/5$, and with pure point-to-point contacts, $c$ can increase to 2/3. To achieve a higher $c$ in the range of $0.7 <c< 0.9$,  it is required to have a $z > 4$, with specific configurations. Therefore, the probability is rather low. This suggests that highly clustered local structures are achieved through a combination of point-to-point and side-to-side contacts rather than maximizing lateral surface contacts between rods. The observed populations at $c = 1$ represent a special case, where the particles are mainly with a $z$ of 2 or 3, with an even (50-50) contact type distribution (Fig.~\ref{fig:microstructure}d). This implies that additional particle contacts tend to form with sparse, point-to-point neighbors rather than within side-to-side connected clusters. The result can be a network where small bundles of side-to-side contacts are connected sparsely, with bundle size directly related to the side contact probability.

\subsection{Rheological measurements}

\begin{figure}[tbp]
    \centering
      \includegraphics[width=0.98\textwidth]{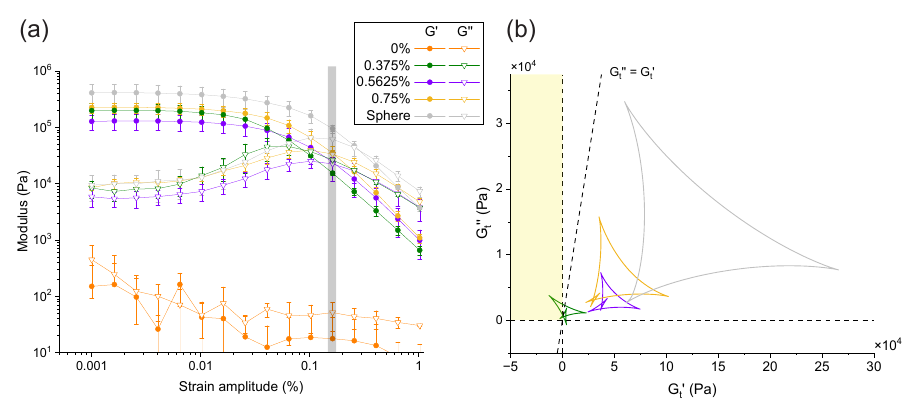}
      \caption[Rheological properties of rod-based capillary suspensions during amplitude sweeps.]{(a) Amplitude sweep of rod-based capillary suspensions with different amounts of secondary liquid. The sphere microparticle-based capillary suspension is made with the protocol from the previous research by Liu~et~al~\cite{Liu2024}, with 10~$\mathrm{\mu}$m microparticles of 15~vol\% and secondary liquid of 0.75~vol\%. (b) The Cole-Cole plot based on the Sequence of Physical Processes of $\gamma_0 = 0.159\% $ is presented with instantaneous loss modulus $G^{\prime\prime}_t$ over instantaneous storage modulus $G^{\prime}_t$. The dashed line represents $G^{\prime}_t = 0$, $G^{\prime\prime}_t = 0$ and $G^{\prime}_t = G^{\prime\prime}_t$, with yellow highlighting the negative $G^{\prime}_t$ and positive $G^{\prime\prime}_t$ region.}
      \label{fig:AmpSweep}
    \end{figure}

Understanding the connection between microstructure and deformation is crucial for designing and optimizing capillary suspensions with anisotropic particles, as the orientation of the particles can greatly influence the rheological properties and functionality of the resulting materials. As shown in Fig.~S2, all of the samples are frequency independent. The moduli as a function of strain amplitude $\gamma_0$ (Fig.~\ref{fig:AmpSweep}a) 
reveals distinct rheological behaviors across different secondary fluid contents, particularly in comparison with pure suspension or capillary suspension using spherical particles. Rod-based systems with secondary fluid exhibit relatively short linear viscoelastic (LVE) regions, maintaining constant moduli only up to approximately 0.01\% strain amplitude, beyond which both $G^{\prime}$ and $G^{\prime\prime}$ begin to deviate from linearity. In contrast, the spherical particle system ($\phi_\mathrm{solid} = 15\% $, $\phi_\mathrm{sec} = 0.75\% $) shows an extended LVE region up to about 0.256\%  strain, suggesting a more robust network structure that can accommodate larger deformations without structural breakdown, possibly due to their isotropy and ability to rebuild bridges~\cite{Liu2024}. The suspension system shows negligible viscoelastic behavior, with nearly immeasurable moduli. The storage modulus in the LVE $G^{\prime}_{LVE}$ appears to be appears to be slightly lower (on the edge of the error bars) for the  0.5625~vol\% sample in both amplitude (Fig.~\ref{fig:AmpSweep}a) and frequency (Supplementary Fig.S2) sweeps. This could be related to the increase in side-to-side contact bundles (Fig.~\ref{fig:microstructure}c), but a general decrease in clustering coefficient (Fig.~\ref{fig:microstructure}e) as the secondary liquid volume fraction increases from 0.375\%  to 0.5625\%. Hsiao~et~al.~\cite{Hsiao2012} have demonstrated that the force is transmitted through more rigid clusters with higher coordination numbers. In the present rod system, despite increased local bundling, the sparse interconnectivity between bundles can reduce the stress-bearing volume fraction, leading to decreased storage modulus. Meanwhile, the enlarged contact area through side-to-side contact results in a higher resistance to rotation and sliding, extending the LVE region~\cite{Pantina2005}. 

Above the LVE, the moduli cross at the flow point ($G^{\prime} = G^{\prime\prime}$). The flow point occurs at $\gamma_0 \approx 0.1\% $  for the systems with $\phi_\mathrm{sec} = 0.375\% $ (green). This short LVE region and low flow point indicate the sensitivity of the rod-particle capillary network to deformation at low secondary liquid volume~\cite{Allard2022}. Systems with higher amounts of secondary fluid display the characteristic $G^{\prime\prime}$ overshoot near $\gamma_0 \approx 0.16\% $, followed by network breakdown.

The critical strain and $G^{\prime\prime}$ overshoot can be rationalized using the viscoplasticity model proposed by Donley~et~al~\cite{Donley2019}, where the classical loss modulus $G^{\prime\prime}$ is divided into an elastic contribution $G^{\prime\prime}_\mathrm{solid}$ and a viscous contribution $G^{\prime\prime}_\mathrm{fluid}$. Within the LVE region, $G^{\prime\prime}_\mathrm{fluid}$, which is associated with the energy dissipation during unrecoverable flow, is negligible because of the nearly full recovery once deformation is ceased. After LVE region, the $G^{\prime\prime}_\mathrm{solid}$, the energy dissipation associated with recoverable deformation decreases drastically due to the collapse of the particle network, at a rate proportional to the decrease of the traditional storage modulus $G^{\prime}$. Meanwhile, $G^{\prime\prime}_\mathrm{fluid}$ increases up with increasing strain. The viscoplasticity value $N_\mathrm{vpf}$, expressed as~\cite{Donley2019},
    \begin{equation}
   N_\mathrm{vpf} = \max \left(\frac{\Delta G^{\prime\prime}_\mathrm{fluid}}{G^{\prime\prime}_\mathrm{fluid}}\right) 
    \end{equation}
quantifies how rapidly a material transitions from recoverable to unrecoverable strain acquisition under deformation. When comparing samples of the same category with similar behaviors in the storage modulus $G^{\prime}$,  $N_{vpf}$ can be simplified as $\max({\Delta G^{\prime\prime}}/{\Delta \gamma})$, the maximum slope of modulus $G^{\prime\prime}$ divided by the strain for values after the LVE and before the $G^{\prime\prime}$ overshoot~\cite{Donley2019, Kamani2021, Kamani2024}, {i.e.} $\gamma_0 =0.01$-- 0.04\%. The corresponding fragility increases with the slope value, implying that the material more quickly acquires unrecoverable deformation. The $G^{\prime}$ slope of the four samples follows the order: 0.375~vol\% ($0.9\times 10^6$ Pa/\%) {\textgreater} sphere ($0.8\times 10^6$ Pa/\%) {\textgreater} 0.5625~vol\% ($0.6\times 10^6$ Pa/\%) = 0.75~vol\% ($0.6\times 10^6$ Pa/\%). At low secondary liquid content (0.375~vol\%), the rod particles predominantly form point-to-point contacts, creating localized stress concentrations, leading to higher viscoplastic fragility as they quickly rupture under strain, causing abrupt structural breakdown. In contrast,  the rods develop more side-to-side contacts and form clusters at higher secondary liquid contents (0.5625 and 0.75~vol\%), distributing external forces over larger contact areas. This results in lower viscoplastic fragility and more gradual yielding behavior~\cite{Kamani2024, Liu2024}. Despite spherical systems having solely point-to-point contact, the higher percolation threshold of spherical particles compared to rod systems results in a more uniform network with intermediate fragility. This behavior illustrates how particle shape influences structural stability under deformation, with rod particles creating networks that are more sensitive to strain compared to their spherical counterparts. resulting in a decreased coordination number, creating a less robust stress-bearing network~\cite{Allard2022}.

Focusing on the sample behavior at $\gamma_0 = 0.16\% $ (gray line in Fig.~\ref{fig:AmpSweep}a), the relationship between the instantaneous loss modulus $G^{\prime\prime}_t$ and storage modulus $G^{\prime}_t$ are plotted in the Cole-Cole plots of Fig.~\ref{fig:AmpSweep}b, where the intracycle behavior is derived from the raw stress and strain responses~\cite{Lee2017}. The capillary suspension with spherical particles demonstrates higher values in both moduli (the trace extends over a larger area) and, with exception of the rod capillary suspension with 0.375\%  secondary liquid (green), all of the samples exhibit elasticity-dominated behavior over the entire cycle, {i.e.} they sit to the right of the dashed line representing  $G^{\prime\prime}_t = G^{\prime}_t$. The sample with 0.375\%  secondary liquid (green) exhibits an intercycle yielding ($G^{\prime\prime}_t \geq G^{\prime}_t$) and indeed even $G^{\prime}_t < 0$ (light yellow shading), indicating a decrease in stress with increasing strain, usually associated with sample recovery after structural breakdown. This negative $G^{\prime}_t$ further confirms the higher viscoplasticity of the low secondary liquid content system, where the rapid transition from solid-like to fluid-like behavior during the oscillation leads to more pronounced structural breakdown and recovery within a single oscillation cycle~\cite{Allard2022, Lee2017, Rogers2017}.



\begin{figure}[tbp]
    \centering
      \includegraphics[width=0.8\textwidth]{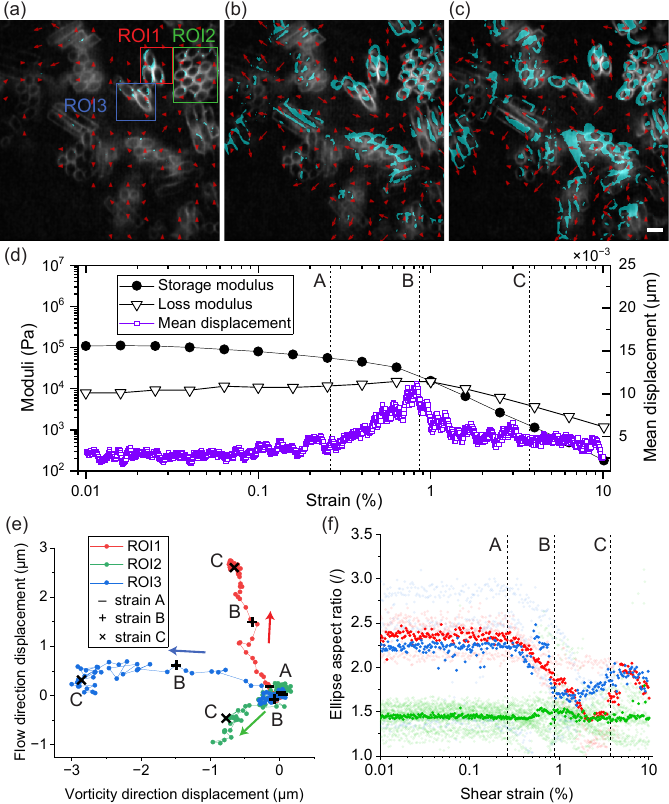}
      \caption[Microscopic flow behavior of clusters during yielding of rod-based capillary suspensions.]{An amplitude sweep was performed on the rheoconfocal microscopy with an gap of 250~$\mathrm{\mu}$m, the image plane was at 36~$\mathrm{\mu}$m. Three images were chosen with positions corresponding to the shear strain amplitudes of $\gamma_0 =  0.26$, 0.89 and 3.72\% . (a-c) Each image is superimposed on the previous one, the differences are shown in cyan pixels and the displacement vectors $\Delta \vec{r}$ are represented by the red arrows, indicating the local displacement within a box of 80 pixel (21.6~~$\mathrm{\mu}$m) square. (d) the shear moduli and the mean displacement are plotted over the strain. The displacement in both flow and vorticity direction, and average degreeularity of each particle of the clusters in ROI1 (red),2 (green), and 3 (blue) are as shown in (e) and (f), respectively, at timestamp strain A (dash), B (plus), and C (cross). }
      \label{fig:Rheoconfocal_all}
    \end{figure}

Rheological measurements alone do not fully elucidate the microstructural response of the system. Therefore, capillary suspension with 0.75~vol\% secondary liquid was chosen to perform rheoconfocal measurements due to its similar structure in comparison to the 0.5625~vol\% sample but with an elevated average elastic modulus. An amplitude sweep was performed at the frequency of 10 rad/s with an image plane at 36~$\mathrm{\mu}$m of the total gap of 250~$\mathrm{\mu}$m. The corresponding visual, rheological, and analyses are plotted in Fig.~\ref{fig:Rheoconfocal_all}. 
Three timestamps (A, B, C) are chosen with positions corresponding to the shear strain amplitudes of $\gamma_0 =  0.26$, 0.89 and 3.72\%, the differences between the images at each strain are shown in Fig.~\ref{fig:Rheoconfocal_all}a-c (where $\gamma_0 = 0.26\% $ is compared to $\gamma_0 = 0\% $) in cyan with the local velocity vectors marked. The rheological measurement and mean displacement are plotted in Fig.~\ref{fig:Rheoconfocal_all}d, with strains from Fig.~\ref{fig:Rheoconfocal_all}a-c marked correspondingly. Before a strain amplitude of $\gamma_0 = 0.26\% $, marked as point a, the mean deformation in Fig.~\ref{fig:Rheoconfocal_all}d is negligible, i.e. within the detection limits, and three clusters (ROI1-3) are observed with slight movements in different directions Fig.~\ref{fig:Rheoconfocal_all}e, as plotted in the flow and vorticity directions. Due to the directionality of anisotropic particles, the displacement can also involve tumbling, which is retrieved from a 2d video of the particle using the mean ellipse aspect ratio (eAR) of the elliptically fitted particle cross-sections, as shown in Fig~\ref{fig:Rheoconfocal_all}f. 
Since eAR values are inversely related to polar angles, particles with an eAR of unity indicate $\varphi$ = 90$^{\degree}$,  with $\varphi$ decreasing with increasing eAR. As shown in Fig~\ref{fig:Rheoconfocal_all}f, the average eAR for each ROI differs from each other, but the eAR for each particle in the cluster is highly uniform (light points), as expected for side-to-side contacts.

The end of the LVE region ($\gamma_0 =0.26$--0.89\% , between points a and b) marks the onset of particle rearrangement. This region typically indicates that the elastic deformation becomes unsustainable and the network starts to rupture, allowing relative movement between particles~\cite{Nelson2019}. This is visually proven by the enhanced mean displacement (Fig.~\ref{fig:Rheoconfocal_all}d), movement of clusters in ROI1 and 3 in flow and vorticity directions respectively (Fig.~\ref{fig:Rheoconfocal_all}e), and the drop in their average eAR (Fig.~\ref{fig:Rheoconfocal_all}f), indicating particle tumbling in addition to displacement (Fig.~\ref{fig:Rheoconfocal_all}e). Interestingly, the three particle clusters all move in different directions -- ROI1 moves in the flow direction and rotates from a higher eAR to lower (decreasing $\varphi$), ROI2 remains in the same position and exhibits only a slight increase in the eAR, and ROI3 displaces in the negative vorticity direction with the eAR decreasing on average with high variability. 

At a strain amplitude of $\gamma_0 = 0.89\% $ (point b), the mean displacement reaches a maximum, likely denoting the limit of elastic deformation where network is stretched to their maximum reversible configuration before bonds begin to break. Shortly after, the flow point is reached, where the network starts to fluidize with majorly irreversible plastic flow~\cite{Bonn2017}, and the network is expected to phase separate into dense structure and expelled bulk liquid, as reported~\cite{Yamamoto2023}. During this process, the bonds where the exerted forces are translated from the top plate break, resulting in decreased mean displacement~\cite{Mueller2009}. This flow point in Fig.~\ref{fig:Rheoconfocal_all}d ($\gamma_0 = 1\% $), is prolonged in comparison to that shown in Fig.~\ref{fig:AmpSweep}, potentially due to a combined effect of enhanced wall slip due to smooth bottom plate (glass for observation)~\cite{Yamamoto2023,Mueller2009,Walls2003} and the increase in side-to-side contact clusters~\cite{Kamani2024,Donley2019} due to compression caused by rheoconfocal thin film loading technique. Post-yielding, e.g. point c, the apparent strain recorded by the rheometer differs from the real strain in the sample due to the slip layer near the top and bottom plates~\cite{Bertola2003}. During this stage, the average displacement of the sample gradually returns to zero. The network structure is disrupted and capillary bridges break, as indicated by cyan in Fig.~\ref{fig:Rheoconfocal_all}c.  The density of particles in this shear plane near the bottom of the plate increases with some of the particles perpendicular to the shear gradient (cyan-marked rectangular particles with high $\varphi$). Accordingly, the eAR values of the clusters fluctuate from shortly prior to point b to after point c, consistent with particle rotation. These three clusters each move randomly, without a clear correlation to the flow direction, with the cluster in ROI2 remaining relatively stationary.

\section{Conclusion}

The study of rod-based capillary suspensions reveals the intriguing interplay between secondary liquid content, particle interactions, and the corresponding microstructure. The present paper demonstrates that capillary forces have a profound influence on particle orientation and network formation, greatly decreasing the onset volume fraction of inducing yield stresses~\cite{Mueller2009,Khan2023}. As the secondary liquid volume increases, the contact type transforms from point-to-point contact to side-to-side contact. Meanwhile, the coordination number and clustering are inversely correlated. Unlike spherical particle systems~\cite{Bindgen2020,Allard2022}, the rod-based capillary suspensions exhibit a complex clustering behavior where increased coordination does not simply lead to more uniform, smectic liquid crystal-type local arrangement, which has not been previously shown for capillary suspensions.

By categorizing the rod particle contact type, we offer a novel methodology to design the microstructure of capillary suspension by changing the secondary liquid volume fractions, and the corresponding rheological response can be predicted. Our rheological measurements denote the unique mechanical sensitivity of these anisotropic networks. The shorter linear viscoelastic regions and distinctive deformation responses highlight the critical role of particle contact configuration in determining material properties~\cite{Kao2022}. Rheoconfocal measurement further uncovers the chaotic nature of cluster movement, revealing the nuanced local mechanical responses that are different from our macroscopic observations, highlighting the influence of anisotropy on network stability under deformation~\cite{Das2021,Das2022}

While our study focused on glass microrods as model prolate stiff microparticles, the fundamental principles of capillary bridges can likely extend to a broader range of anisotropic particle systems due to the high contact area between side-to-side contact-type particle pairs. This study also demonstrates how capillary forces, which are significantly higher than van der Waals or other forces studied in literature~\cite{Das2021,Kao2022,Das2022,Shakeel2021,Lang2019,Dhont2006,Calabrese2023}, can fundamentally alter network structures and rheological properties.
Our research paves the way for using rod-based capillary suspensions as versatile precursors for advanced material design. The insights gained with regard to particle orientation, cluster formation, and yielding behavior provide a foundation for developing materials with precisely tunable mechanical properties through controlled anisotropic particle interactions. By controlling the capillary forces between particles through secondary liquid volume fraction, we can create materials with desired complexity and functionality, with applications ranging from 3D printing ceramics to novel food formulations. 

In future work, we aim to expand the study to include more industry-relevant anisotropic particles such as starch granules, protein aggregates, and ceramic particles, where particle shape plays an important role in the capillary suspension formulation. In parallel, we will systematically investigate the percolation threshold and its corresponding rheological properties. Moreover, modifying particle surface chemistry to alter three-phase contact angles could also provide additional control over the transition between contact types~\cite{Eyley2014}, enabling precise tuning of mechanical properties.

\section*{Author Contributions}

\textbf{Lingyue Liu}: Conceptualization, Methodology, Software, Investigation, Writing – original draft. \textbf{Sebastian Gassenmeier}: Methodology, Software. \textbf{Erin Koos}: Funding acquisition, Conceptualization, Methodology, Writing – review \& editing, Supervision.

\section*{Declaration of competing interest}

The authors declare that they have no known competing financial interests or personal relationships that could have appeared to influence the work reported in this paper.

\section*{Acknowledgements}
The authors would like to thank financial support from the European Union’s Horizon 2020 research and innovation programme under the Marie Sk\l{}odowska-Curie grant agreement No 955612 and International Fine Particle Research Institute (IFPRI). We would like to thank Minne Paul (Pavlik) Lettinga for his support and invaluable comments.

\section*{Data availability}
Data will be available on request.



%

\clearpage
\onecolumngrid
\appendix

\renewcommand{\thesection}{S\arabic{section}}
\renewcommand{\thetable}{S\arabic{table}}
\renewcommand{\thefigure}{S\arabic{figure}}
\setcounter{figure}{0}
\setcounter{table}{0}
\setcounter{page}{1}
\renewcommand{\floatpagefraction}{.9}%

\section*{Supplementary material}

\begin{figure}[h!]
\centering
  \includegraphics[width=0.5\textwidth]{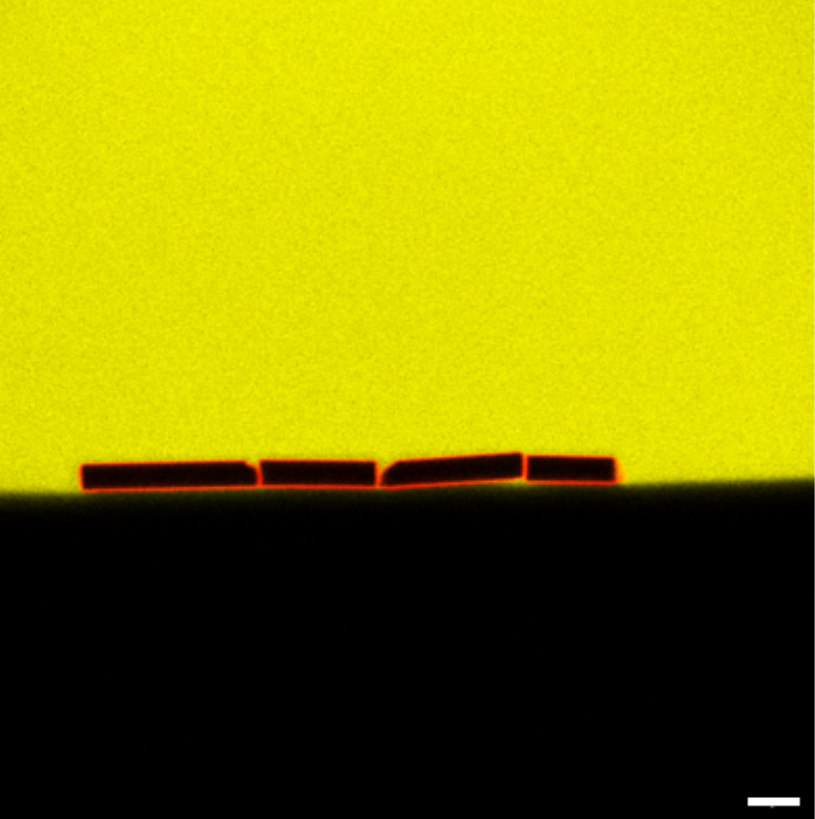}
  \caption[Contact angle measurement of rod particles at liquid-liquid interface.]{Contact angle measurement of rod particles at bulk (black) and secondary liquid (yellow) interfaces. The scale bar is 10~$\mathrm{\mu}$m.}
  \label{fig:sup:CA}
\end{figure}

\begin{figure}[h!]
\centering
  \includegraphics[width=0.6\textwidth]{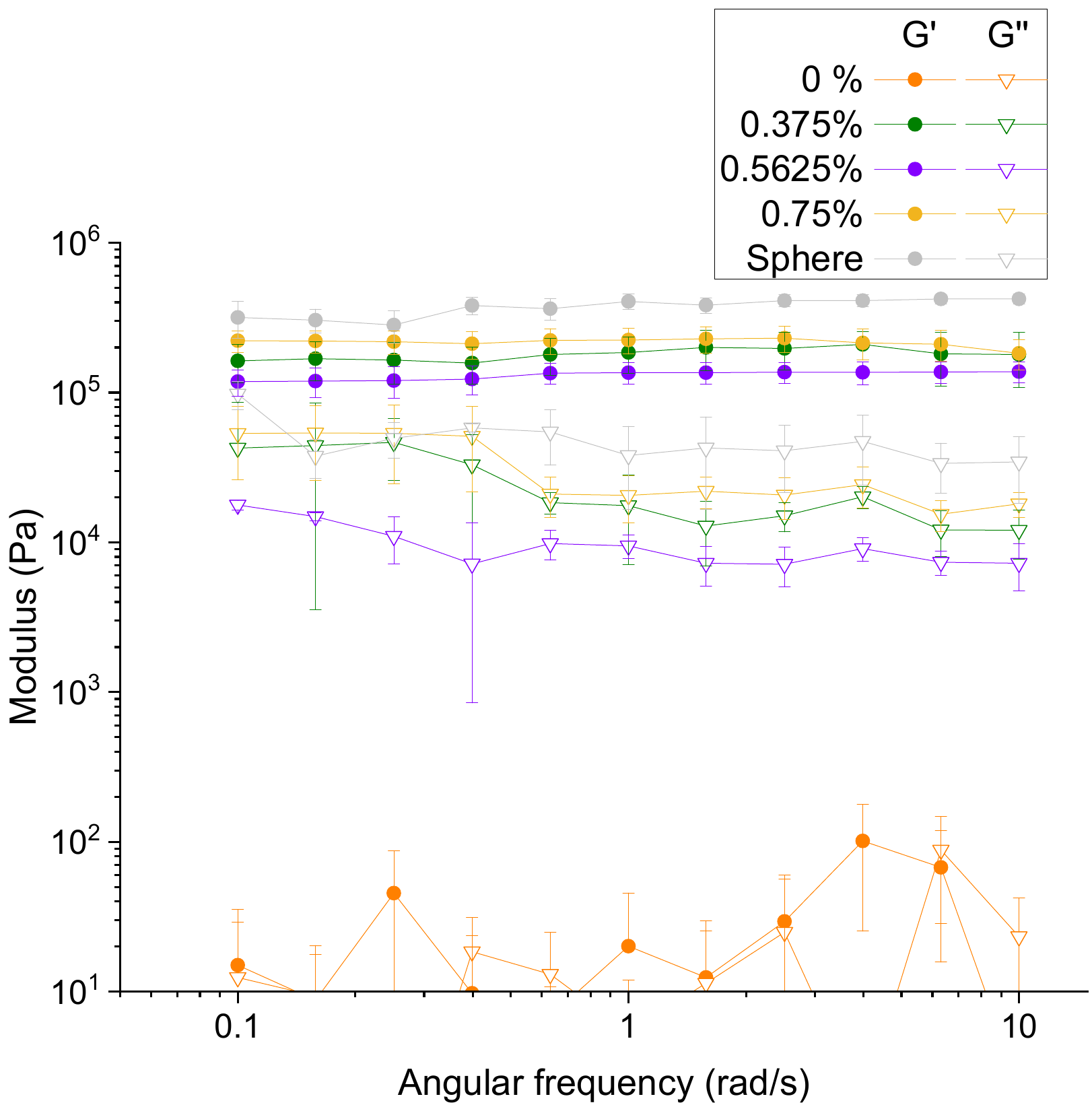}
  \caption[Frequency sweeps of rod- and sphere-based capillary suspensions.]{Frequency sweep of rod-based capillary suspensions with different amounts of secondary liquid. The sphere microparticle-based capillary suspension is made with 15~vol\% 10~$\mathrm{\mu}$m microparticles and 0.75~vol\% secondary liquid.}
  \label{fig:sup:fsweep_flowcurve}
\end{figure}

\begin{figure}[h!]
\centering
  \includegraphics[width=0.3\textwidth]{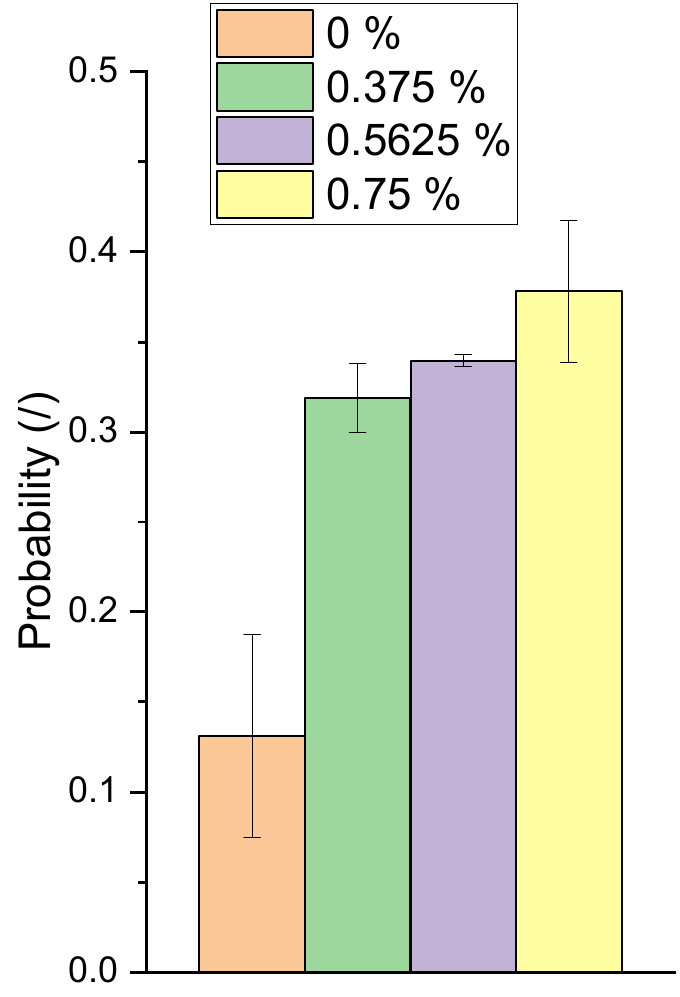}
  \caption[Averaged side contact probability with different amounts of secondary liquids.]{Averaged side contact probability with different amounts of secondary liquids.}
  \label{fig:sup:AverageSideContact_secliquid}
\end{figure}

\begin{figure}[h!]
\centering
  \includegraphics[width=0.7\textwidth]{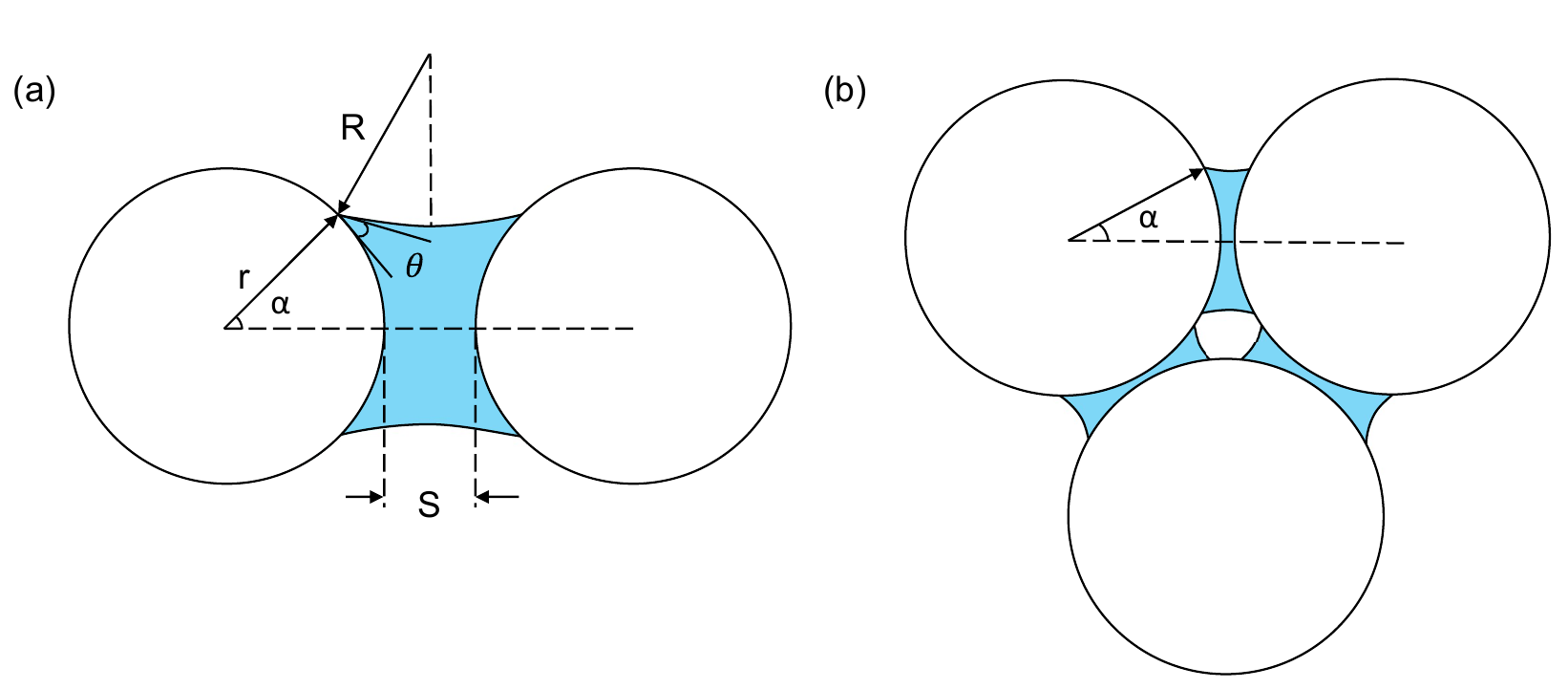}
  \caption[Schematic representation of liquid bridges between rod particles.]{Cross section of a liquid bridge between (a) two (b) three parallel rods in the central region. Schematics showing key parameters: r (rod radius), R (bridge curvature radius), $\theta$ (contact angle), S (separation distance), and $\alpha$ is the angle between the line connecting the centers of the cylinders and the radius to the liquid-liquid boundary.}
  \label{fig:sup:rods_illustration}
\end{figure}

\begin{figure}[h!]
    \centering
      \includegraphics[width=0.6\textwidth]{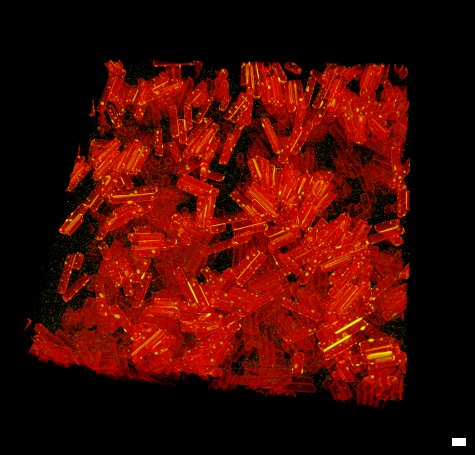}
      \caption[3D reconstructed confocal micrographs of capillary suspension.]{3D reconstructed confocal micrographs of capillary suspension with a secondary liquid of 0.75\% ($\phi_\text{sec}/\phi_\text{solid}$ = 5\%), the scale bar is 10~$\mathrm{\mu}$m. Red regions show rod particles (shells), while yellow areas represent capillary bridges, and the dark background is the bulk liquid.}
      \label{fig:sup:3dBridge}
    \end{figure}

\begin{figure}[h!]
\centering
  \includegraphics[width=0.9\textwidth]{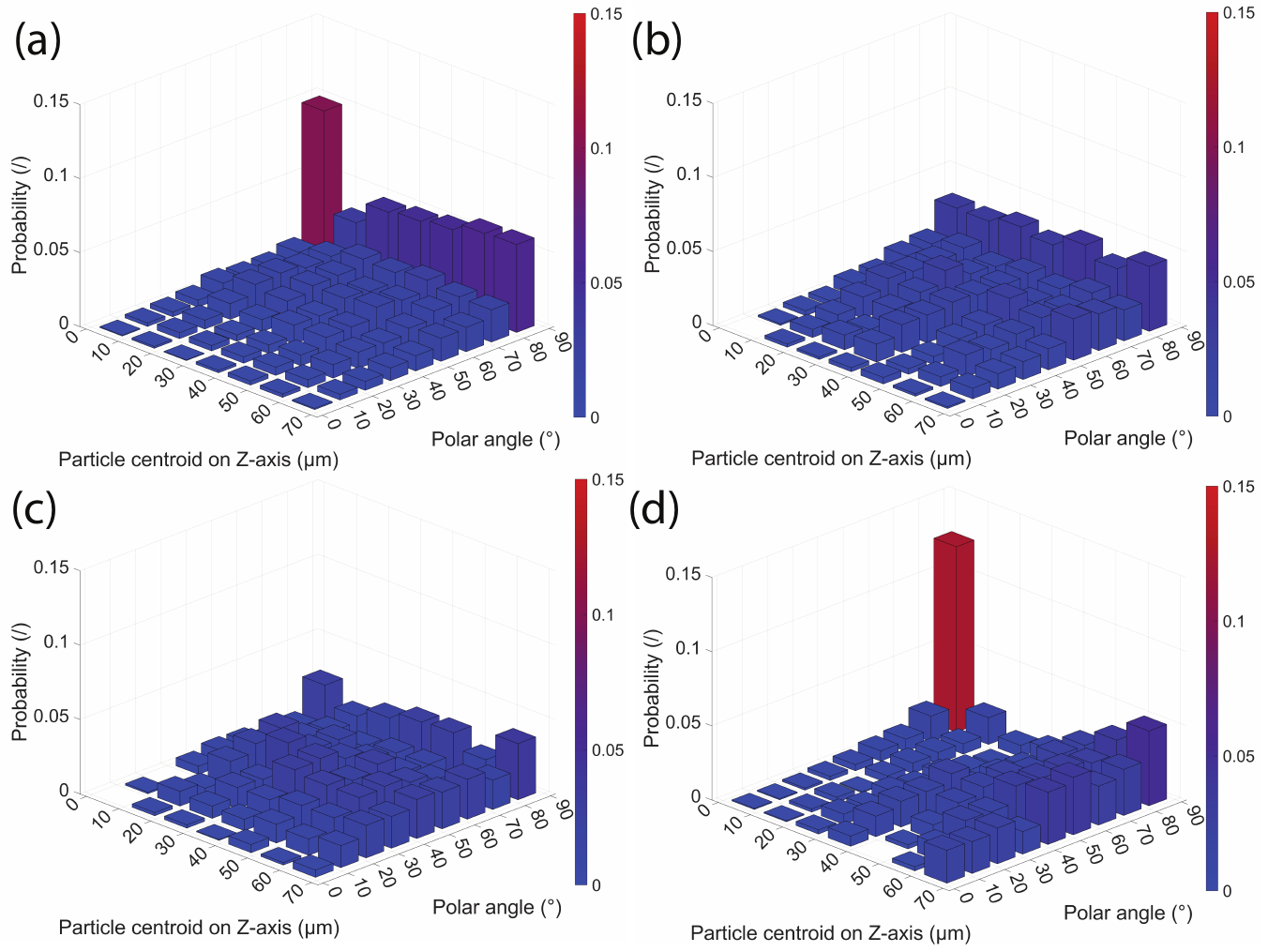}
  \caption[Polar angle probability as a function of particle position in capillary suspensions with different amounts of secondary liquid.]{Polar angle probability as a function of particle position in capillary suspensions with different amounts of secondary liquid (a) 0\% ($\phi_\text{sec}/\phi_\text{solid}$ = 0.0\%) (b) 0.375\% ($\phi_\text{sec}/\phi_\text{solid}$ = 2.5\%) (c) 0.5625\% ($\phi_\text{sec}/\phi_\text{solid}$ = 3.75\%) (d) 0.75\% ($\phi_\text{sec}/\phi_\text{solid}$ = 5\%).}
  \label{fig:sup:PolarAngleInZ}
\end{figure}

\begin{figure}[h!]
\centering
  \includegraphics[width=0.9\textwidth]{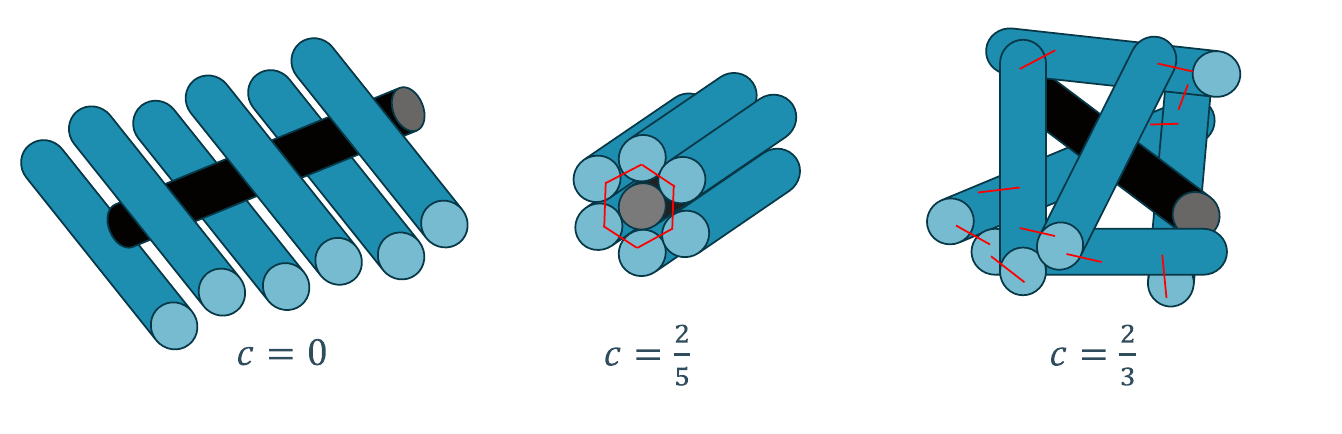}
  \caption[Relationship between rod arrangement and clustering coefficient.]{Schematic showing the clustering coefficient for various configurations of rods with a coordination number of 6 in 3D. Black particle is the center particle with neighboring particles represented in blue. Red lines show the connections between neighboring particles.}
  \label{fig:sup:3D_cluster}
\end{figure}

\end{document}